\documentclass[%
reprint,
superscriptaddress,
 amsmath,amssymb,
 aps,
prb
]{revtex4-2}
\usepackage{graphicx}
\usepackage{dcolumn}
\usepackage{bm}
\usepackage{chemformula}
\usepackage[colorlinks=true,citecolor=blue,linkcolor=blue,urlcolor = blue, breaklinks=true]{hyperref}

\begin{document}


\title{Simple 4-segment thermal cycling pyroelectric measurement protocol for differentiating between ferroelectric and non-ferroelectric materials
}


\author{Aditya A. Wagh}
\email[]{adityawagh@iisc.ac.in}
\affiliation{Department of Physics, Indian Institute of Science, Bangalore 560012, INDIA}

\author{Shwetha G. Bhat}
\email[]{shwetha@iisc.ac.in}
\affiliation{Department of Physics, Indian Institute of Science, Bangalore 560012, INDIA}

\author{V. K. Anusree}
\affiliation{Functional Oxides Research Group, Department of Physics, Indian Institute of Technology, Madras 600036, INDIA}

\author{P. N. Santhosh}
\affiliation{Functional Oxides Research Group, Department of Physics, Indian Institute of Technology, Madras 600036, INDIA}

\author{Suja Elizabeth}
\affiliation{Department of Physics, Indian Institute of Science, Bangalore 560012, INDIA}

\author{P. S. Anil Kumar}
\affiliation{Department of Physics, Indian Institute of Science, Bangalore 560012, INDIA}

\begin{abstract}
The rare-earth chromates (RECrO$_3$) and manganites (REMnO$_3$) where, RE = Eu, Y, Dy, Ho, Gd are constantly under scrutiny in search of room temperature magnetoelectric multiferroics. However, the artefacts and undesirable signal in some of the measurements pose a severe challenge in confirming the ferroelectric (FE) phase, especially in reference to pyroelectric current measurement technique. In this regard, we propose a simple modified approach to pyroelectric current measurement named as \textit{4-segment thermal cycling protocol}. This protocol assists in isolating the elusive, irreversible thermally stimulated current from the currents associated with spontaneous and reversible nature of the electric polarization in FE phase. In order to explain working principle of the protocol, we have compared simulated response of two hypothetical materials; an FE material free of space charges and a paraelectric material possessing only space charges. Further, we experimentally verify these new protocols in a single crystal of prototype ferroelectric material, Glycine Phosphite. This report primarily focuses on detailed investigation of ferroelectricity using the proposed protocol in two polycrystalline materials, HoCrO$_3$ and DyFe$_{0.5}$Mn$_{0.5}$O$_3$ where, the former has been reported to be multiferroic earlier. Our elaborative and careful approach to pyroelectric studies expound on the absence of reversible spontaneous electric polarization at temperature ranges tested in both, HoCrO$_3$ and DyFe$_{0.5}$Mn$_{0.5}$O$_3$.
\end{abstract}


\maketitle

\section{Introduction}

There are materials in which ferroelectric (FE) and magnetic ordering coexist are called multiferroics \cite{Tokura2007,Tokunaga2008,Tokunaga2009,Kitagawa2010,Fiebig2002,Yamasaki2007}. The theoretical predictions of magnetoelctric (ME) property in orthochromates, orthoferrites and manganite perovskite materials \cite{Yamaguchi1973,Zvezdin2008,Fennie2008,Iusan2013} along with magnetic origin of strong ME multiferroicity in REMnO$_3$ (RE = Tb, Dy, Gd) systems with spiral order \cite{Kimura2003,Kimura2005} paved a way to search for new room-temperature ME multiferroics. Later, plethora of other materials belonging to REMnO$_3$, REFeO$_3$, RECrO$_3$, RE$_2$B$_2$O$_6$ (RE = Rare-earth cation, B = 3\textit{d}-magnetic cation) and spiral systems such as hexaferrites are under constant consideration for studies over last two decades \cite{Feng2010,Chai2012,Azuma2005,Lee2011,Lee2011a,Rajeswaran2012,Kitagawa2010,
Kimura2003,Wang2003,Tokunaga2008,Tokunaga2009}. 

A key aspect of experimental research in multiferroics is about confirming the presence of FE ordering or, in other words, spontaneous electric polarization ($P_{S}$) i.e. switchable by an electric field ($E$). For instance, the coexistence of FE and magnetic phases were experimentally realized in YMnO$_3$ system using optical second harmonic generation \cite{Fiebig2002}. On the other hand, most of the electric transport experiments utilize the fact that the variation of temperature ($T$), magnetic field ($H$) or $E$ (in case of ME multiferroics) alters volume bound charges (i.e. $P_{S}$). This change in bound charges ($\Delta P_{S}$) manifests itself in measurable compensation currents ($I_{C}$), if electrodes on two opposite faces of the sample are shorted. The P-E loop tracer method records $I_{C}$ while scanning the external $E$ to high magnitudes. Under such fields in lossy materials, currents constituting free-carriers pose serious problems in the reliable estimation of $\Delta P_{S}$. Therefore, another approach of tracking $I_{C}$ while scanning the $T$ is widely used \cite{Kimura2003,Tokunaga2008,Yamasaki2007}, popularly known as pyroelectric current ($I_{P}$) measurement. It is mainly due to the ease of performing experiments in-house without using sophisticated setups such as Synchrotron sources \cite{Sakai2011}. Notably, an absence of external $E$ during the signal-measurements helps in avoiding possible spurious contributions due to free-carrier currents. This linear $T$-ramp method or Byer-Roundy method \cite{Byer1972}, has been widely adopted in many studies \cite{Lee2011,Rajeswaran2012,Meher2014,Ghosh2014,Ghosh2015,
Hao2016,Sharma2016,Indra2016,Xie2017,Dey2019,Chatterjee2019,Su2015} in the past. In addition, laser assisted heating techniques are also an alternative for $I_{P}$ vs. $T$ measurements. Developments such as periodic temperature change (PTC) in the optical techniques \cite{chynoweth1956dynamic,holeman1972sinusoidally} are also widely utilized, apart from linear $T$-ramp methods.

In a typical $I_{P}$ measurement, in order to achieve single electric domain, the sample is first poled with $E$ in the paraelectric (PE) phase and then cooled to the FE phase, after which the poling field is removed. As a precaution, hours-long waiting is done for depolarization of induced space charges. It is significant to highlight the importance of time-scales associated with this depolarization, often overlooked in experiments. The depolarization time-scales are generally high at low $T$ and consequently, the hours-long waiting time would not release all trapped space charges (as discussed in Appendix). Besides, as depolarization time-scales drop rapidly when $T$ increases, the depolarization current generated due to induced space charge polarization ($P_{Induced}$) interferes with the current due to depolarization of $P_{S}$. These unwanted thermally stimulated currents ($TSC$) deceive us mainly in two ways; (1) A broad peak in $TSC$ during warming with linear ramp rate can be misinterpreted and assigned to FE-PE transition. (2) Pre-poling with the reversed direction of $E$, results in reversing the sign of $TSC$. Therefore, the above points can be precarious criteria for confirming FE. In this line of thought, some studies have proved the absence of multiferroicity in the material, despite the estimated polarization from $I_{P}$ measurements yielded non-zero values \cite{Nhalil2015,Blasco2017,Yang2020,Anusree2020}.

The original Byer-Roundy method has been modified numerous times in the past. Notably, PTC methods that employed On/Off sequence of pulsed LASER \cite{chynoweth1956dynamic} and the sinusoidal modulation of temperature using LASER beam \cite{holeman1972sinusoidally} are popular optical methods. Even though, the PTC techniques with reference to $I_{P}$ measurements are well established optical methods, adapting it for measurements at high-magnetic fields inside low-temperature cryostats warrants some sophistication and hence are less popular in multiferroics community. Here, in this context, the PTC techniques that employ small temperature modulation using resistive heating can offer great flexibility. For instance, earlier reports \cite{Garn1982,Sharp1982} showed that the two depolarization current contributions in $I_{C}$, originated from both $P_{S}$ and $P_{Induced}$, can be separated by employing a single low-frequency sinusoidal $T$-modulation over a linear ramp. However, generation of such precise single-frequency modulation essential for the experiment could be challenging in some laboratory setups. To overcome difficulties mentioned in the above techniques, we propose here a simple modified protocol which is employable with resistive heating. We call this protocol a \textit{4-segment thermal cycling protocol}. The protocol exploits the fact that the signal $I_{P}$ is proportional to thermal ramp rate and employs thermal cycling comprised of discrete thermal ramp rates (as discussed in Section \ref{pyrosimulation}.). However, it is qualitatively different from the PTC techniques. We have employed this protocol and simulated the response (induced $I_{C}$) in two hypothetical (ideal) systems; 1) Ideal FE material having no space-charges, and 2) PE material possessing only induced space charges. Further, we have extensively analyzed the existing theory of $TSC$ in a non-FE material revealing different mechanisms responsible for observing sizeable amounts of current in $I_{P}$ measurements. The proposed thermal cycling protocol was experimentally verified and shown in a prototype of standard FE material, Glycine Phosphite single crystal (discussed in Section \ref{GPI and GMO}). Additionally, the relevance of repetitive field cycling in ME current measurements was demonstrated in the case of ME multiferroic GdMnO$_3$. Researchers had earlier reported the existence of multiferroicity in the perovskite material HoCrO$_3$ \cite{Ghosh2015, Saha2014}. However, our preliminary investigations did not yield evidence of FE in the range of temperatures studied. Another family of materials of interest is the B-site disordered ABO$_3$ systems, where renewed interest of multiferroicity in the family \cite{yuan2016observation,rajeswaran2012ferroelectricity} prompted us to choose the DyFe$_{0.5}$Mn$_{0.5}$O$_3$ compound for a property-specific investigation. This study details the comparison of experimental data of both materials, HoCrO$_3$ and DyFe$_{0.5}$Mn$_{0.5}$O$_3$, and initial results of the application of the afore-mentioned \textit{4-segment thermal cycling protocol} for differentiating the spontaneous and induced charges. Further, it can help to interpret data realistically and identify ferroelectric materials unambiguously in future studies.

\begin{figure*}
\centerline{\includegraphics[width=16cm,keepaspectratio]{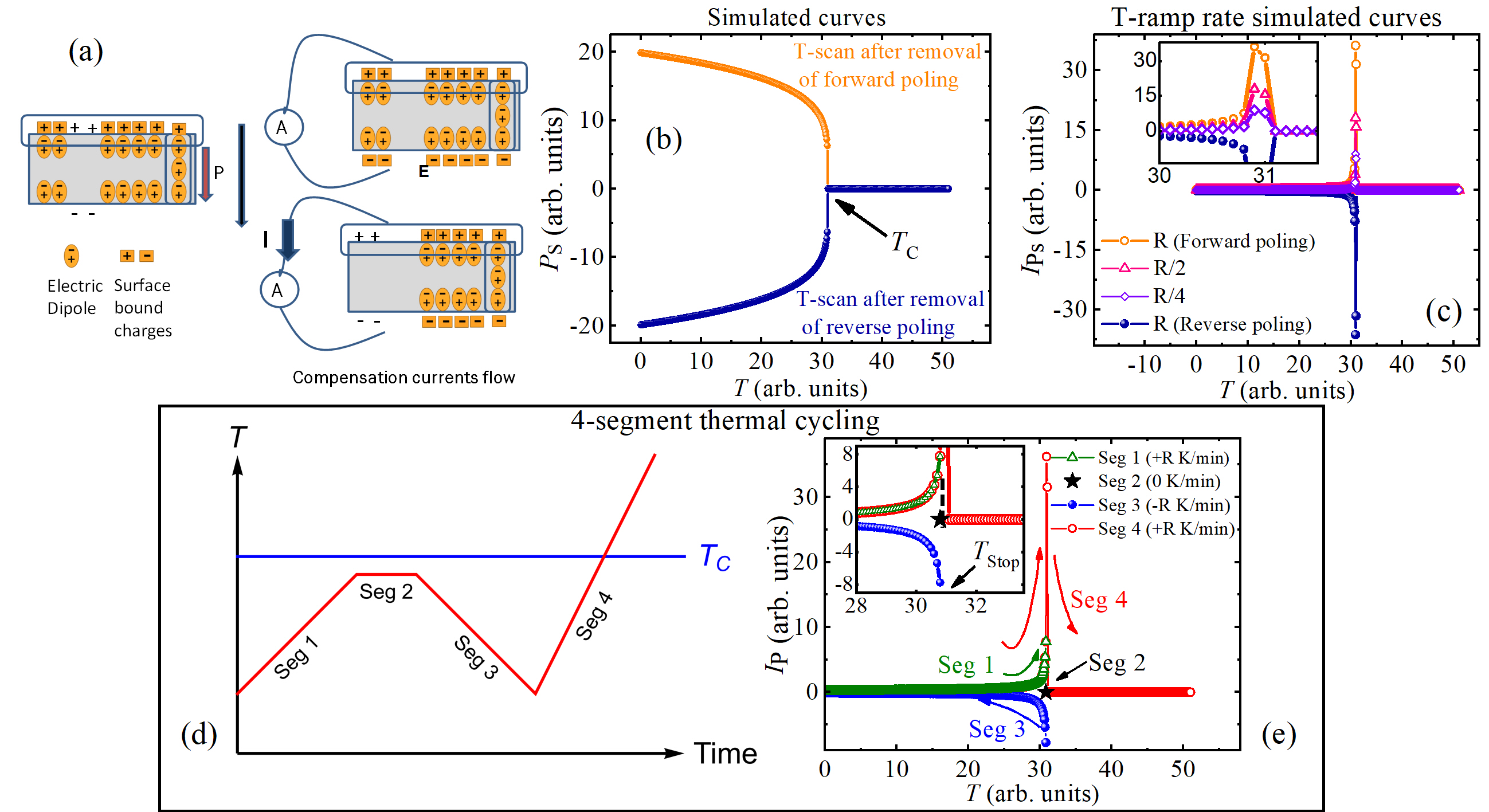}}
\caption{(a) Schematics of the pyroelectric current ($I_{P}$) measurements for a typical ferroelectric material. (b) The expected variation of $P_{S}$ with $T$ for forward and reverse poling scenarios. (c) $I_{P_{S}}$ vs. $T$ as a function of different ramp rate in arbitrary units. Inset shows a zoom-in view near the ferroelectric $T_{C}$. (d) \textit{4-segment thermal cycling protocol} below the ferroelectric $T_{C}$ of the material, represented as different segments in increasing order. Inset of (d) shows a zoom-in view of the $I_{P_{S}}$ near the $T_{Stop}$ (where, $T_{Stop} < T_{C}$) for different segments of the protocol.}
\label{fig1}
\end{figure*}

\section{Thermal cycling of depolarization currents: Spontaneous Electric Polarization vs. Induced Polarization}
\label{pyrosimulation}

In this section, we discuss the \textit{4-segment thermal cycling protocol} assisting in identifying artifacts in $I_{P}$ measurements. We consider two individual materials with ideal behaviour undergoing a typical poling process: (1) FE material consisting of pure $P_{S}$, and (2) non-FE material comprising only $P_{Induced}$. For both these cases, we compare and discuss simulated depolarization currents resulting as a response to our \textit{4-segment thermal cycling protocol}.

\subsection{Depolarization of spontaneous polarization}
\label{pyro}

FE materials possess $P_{S}$ which is essentially switchable by $E$. Typically, before $I_{P}$ measurements, the samples are poled much above $T_{C}$ and cooled down to well inside the FE phase to attain a single domain. Subsequently, the poling field is removed, and the sample is shorted (to discharge the $P_{Induced}$) before the experiments. Figure \ref{fig1}(a) shows that the generated $P_{S}$ is denoted by electric dipoles forming head-to-tail chains in the bulk of the material and localizing electrons in the electrode as surface-bound charges.

Notably, as the sample is warmed, thermal energy randomizes the dipole moments and the net $P_{S}$ decreases with $T$ (see Fig. \ref{fig1}(b)). Typically, the variation of $P_{S}$ with $T$ is reversible.  Fig. \ref{fig1}(b) shows a typical variation of $P_{S}$ with $T$ across $T_{C}$ when the sample is pre-poled in forward or reverse direction.

As the sample is warmed, the polarization of the sample decreases with an accompanying decrease in dipoles per unit volume in the bulk and consequently, It releases few bound surface charges to both electrodes. If the electrodes are shorted by a conducting wire then, $I_{C}$ flows in the wire. The resulting  $I_{C}$ flow is associated with the change in polarization. Thus, the induced pyroelectric (compensation) current is directly proportional to time rate of change of $P_{S}$ (see Eq. (\ref{eq:1})) with the proportionality constant, termed pyroelectric coefficient, $C$.

\begin{equation}
\label{eq:1}
I_{P_{S}} = C.\frac{dP_{S}}{dt}
\end{equation}
\begin{equation}
\label{eq:2}
I_{P_{S}} = C.\frac{\partial{P_{S}}}{\partial{T}}.\frac{\partial{T}}{\partial{t}}
\end{equation}

$\frac{dP_{S}}{dt}$ in Eq. (\ref{eq:1}) can be written as product of two partial derivatives as in Eq. (\ref{eq:2}). In this case, sign and magnitude of the induced pyroelectric current at a particular $T$ is determined by sign and magnitude of both $\frac{\partial{P_{S}}}{\partial{T}}$ and $\frac{\partial{T}}{\partial{t}}$ at that $T$. Forward and reverse poling of the sample result in opposite signs of the $\frac{\partial{P_{S}}}{\partial{T}}$ and as a result, exhibit opposite signs in corresponding $I_{P_{S}}$ (see Fig. \ref{fig1}(b) and (c)). Figure \ref{fig1}(c) also shows that the magnitude of $I_{P_{S}}$ scales linearly with the ramp rate, $\frac{\partial{T}}{\partial{t}}$. As the variation of $P_{S}$ with $T$ is reversible, values of both, $P_{S}$ and $\frac{\partial{P_{S}}}{\partial{T}}$, are unique at a particular $T$ for a certain poling scenario. Therefore, the sign and magnitude of $I_{P_{S}}$ can be directly manipulated by the sign and magnitude of the thermal ramp rate, $\frac{\partial T}{\partial t}$.  
The above relations and the knowledge that variation of $P_{S}$ with temperature is reversible help us formulate a \textit{4-segment thermal cycling protocol}; 

	\textit{Segment 1}: $\frac{\partial T}{\partial t}>0$, warming from an initial temperature ($T_{0}$) to the stopping temperature ($T_{Stop}$) where, $T_{Stop} < T_{C}$.
	
	\textit{Segment 2}: $\frac{\partial T}{\partial t}=0$, halting at the temperature $T_{Stop}$ for a certain time interval, $Time_{Halt}$.
	
	\textit{Segment 3}: $\frac{\partial T}{\partial t}<0$, cooling from $T_{Stop}$ to $T_{0}$.
	
	\textit{Segment 4}: $\frac{\partial T}{\partial t}>0$, warming again from $T_{0}$ to the end temperature ($T_{End}$) where $T_{End}>T_{C}$.

It is evident that $I_{P}$ in the \textit{Segment 1} and the \textit{Segment 3} will yield mirror-like symmetric inverted currents (see Fig. \ref{fig1}(d)). \textit{Segment 2} gives zero $I_{P}$ denoted by a black star symbol in Fig. \ref{fig1}(d). While, the \textit{Segment 4}, over the temperature range, $T_{0}$ to $T_{Stop}$,  yields an identical $I_{P}$ as that of the \textit{Segment 1}. Further, warming results in a peak at a higher $T$ before dropping to zero at $T_{C}$ (see Fig. \ref{fig1}(d)).

Above \textit{4-segment thermal cycling protocol} not only exploits spontaneous and switchable nature of the polarization in FE material but also, carefully tests for essential reversibility of $P_{S}$. Further, there is a prospect of identifying the reversible component of spontaneous polarization (ferroelectricity) and corresponding pyroelectric current, from the non-reversible slow-decaying induced polarization and related $TSC$, respectively, which is given in the following subsection. 

\subsection{Depolarization of induced polarization}
\label{TSC}

In this section, we consider material that possesses only $P_{Induced}$ caused by the typical poling protocols. The theory of the resultant $TSC$ due to $P_{Induced}$ (which is different from the $I_{P}$) is dealt in detail in Appendix. Interestingly, $TSC$ as a function of $T$ can exhibit broad peak features. Hence, it is essential to distinguish the contributions from these non-pyroelectric origins such as $TSC$ from $I_{P}$ in experimental scenario.

When a non-FE material is poled with an electric field ($E_{P}$) at a temperature  ($T_{P}$), $P_{induced}$ develops in the bulk of the material depending on the time interval of applied $E_{P}$. Considering dipolar relaxation phenomena, the depolarization of $P_{induced}$ is dependent on both time ($t$), $T$ and in turn, the temperature ramp rate ($b$) (Refer in Appendix- Eq. (\ref{eq:PdwithTime}), Eq. (\ref{eq:PdwithT}) and Fig. \ref{appendixfig1}). Thus, it is evident that the functional behavior of the $P_{induced}$ is naturally different from that of $P_{S}$. When the poled sample cools to $T_{0}$, the time constant associated with the depolarization of $P_{induced}$ at $T_{0}$ assumes large values depending on the activation energy ($E_{a}$) of the mechanism. Waiting at $T_{0}$ for few hours after removing $E_{P}$ ensures that the non-pyroelectric current reduces to a low value at that $T$, near to the detection limit of experimental measurements. Nonetheless, a cautious approach is advised  when temperature is increased from $T_{0}$ to higher $T$ while recording this current. The time constant related to depolarizing of $P_{induced}$ decreases exponentially, releasing more trapped charges while constituting a very broad peak of $TSC$ at higher temperatures (Refer to Fig. \ref{appendixfig1}(b) in Appendix). These values of $TSC$ (or $J_{TS}(T)$) are simulated using Eq. (\ref{eq:JdwithT}) of Appendix. 

It is noteworthy here that, the position of the broad peak of $TSC$ (as in Fig. \ref{appendixfig1}(b)) shifts to higher $T$ as $b$ is increased, unlike the fixed peak-position of $I_{P_{S}}$ in case of a FE material (as in Fig. \ref{fig1}(c)). Hence, this comparison could be used as one of the criteria to assign the nature of the peak to the pyroelectric/non-pyroelectric contribution. However, the reversibility of the $I_{C}$ due to $P_{S}$ is also another essential factor for distinguishing the $P_{S}$ from $P_{induced}$. Thus, the \textit{4-segment thermal cycling protocol} (as mentioned in Section \ref{pyrosimulation} A) can be utilized in this context of $TSC$ to verify the nature of a slow decay and irreversibility of $P_{induced}$.

\begin{figure}
\centerline{\includegraphics[width=8cm,keepaspectratio]{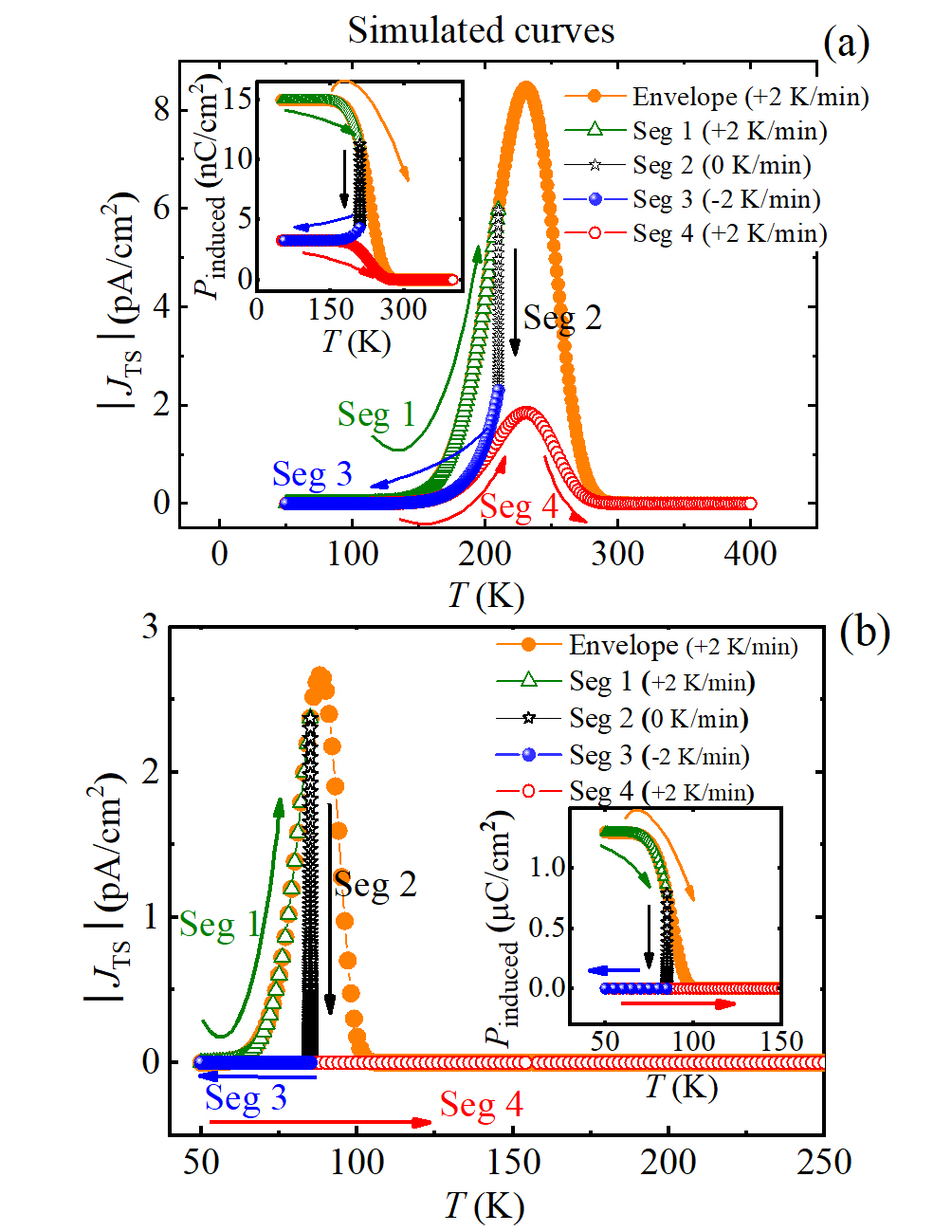}}
\caption{Simulation of thermally stimulated currents following the \textit{4-segment thermal cycling protocol}. (a) $J_{TS}$ as a function of $T$ for a slower mechanism of depolarization and inset is the depolarization of $P_{induced}$ with $T$. (b) $J_{TS}$ vs. $T$ for a relatively faster mechanism of depolarization (Inset: $P_{induced}$ vs. $T$).}
\label{fig2}
\end{figure}

Further, we extend the simulations of $J_{TS}(T)$ to generate the \textit{4-segment thermal cycling protocol} curves with a poling process carried at $T_{P}$ = 400 K and the data are shown in Fig. \ref{fig2}. The inset of the Fig. \ref{fig2} is $P_{induced}$ vs. $T$ estimated from Eq. (\ref{eq:PdwithT}) of Appendix. For simplicity, we consider two possible scenarios of $TSC$, defined by two pairs of $E_{a}$ and time constant ($\tau$) to generate depolarization of $J_{TS}(T)$ or $P_{induced}$, to mimick (1) a slower  mechanism with $E_{a}$= 0.176 eV and $\tau$ = 88 ms (Fig. \ref{fig2}(a)) and, (2) a relatively faster mechanism with $E_{a}$= 0.189 eV and $\tau$ = 4 ns (Fig. \ref{fig2}(b)). The values chosen for $E_{a}$ and $\tau$ are relevant from the experimental context and discussed separately in Section \ref{HCO and DFMO}. The values chosen (common for both mechanisms) for the \textit{4-segment thermal cycling protocol} are as follows: $T_{0}$ = 50 K, $T_{End}$ =  400 K, and $b$ = 2 K/min. Whereas, $T_{Stop}$ = 210 K (85 K) and $Time_{Halt}$ = 30 min (60 min) is considered for slower (faster) mechanism, in Fig. \ref{fig2}(a) (Fig. \ref{fig2}(b)).

The envelope curves in Fig. \ref{fig2} are the data from $T_{0}$ to $T_{End}$ without any halting at intermediate temperature. At first, we will consider the slower mechanism (Fig. \ref{fig2}(a)). At the end of \textit{Segment 2} (at $T_{Stop}$), there is still some non-zero $P_{induced}$ left in the system (see inset of Fig. \ref{fig2}(a)), leaving behind some amount of $J_{TS}(T)$ at $T_{Stop}$. Further, when  $\frac{\partial T}{\partial t}<0$ in \textit{Segment 3}, the $J_{TS}(T)$ continues to drop to even lower values. A non-zero value of $P_{induced}$ is noticed at the beginning of \textit{Segment 4}, which is smaller than the value available at the beginning of \textit{Segment 1} at $T_{0}$ (see inset of Fig. \ref{fig2}(a)). Thus, a smaller magnitude of peak in $J_{TS}(T)$ results while warming from $T_{0}$ to $T_{End}$. However, the faster mechanism, has a different nature of $TSC$ due to a distinct choice of $E_{a}$ and $\tau$ (See Fig. \ref{fig2}(b)). Here, unlike the \textit{Segment 2} of the slower mechanism, the $J_{TS}(T)$ drops to zero while halting at $T_{Stop}$, resulting in a complete depolarization of $P_{induced}$ at the end of \textit{Segment 2} itself, as seen in the inset of Fig. \ref{fig2}(b). Consequently, the value of $P_{induced}$ in \textit{Segment 3} and \textit{Segment 4} continues to be at zero. Ergo, the analysis and the comparison of two scenarios of dipolar relaxation with a distinct choice of $E_{a}$ and $\tau$ adapted in \textit{4-segment thermal cycling protocol} represents the depolarization of $P_{induced}$ in two distinct cases. However, there are other mechanisms of $TSC$ generation that have not been under consideration here such as, injected charges in the vicinity of the electrode/material interface due to poling. But, the overall behavior of the $TSC$ response is the same as that in our study.

The most salient feature here is the comparison of simulated results of depolarization of $P_{S}$ and $P_{induced}$ in the context of \textit{4-segment thermal cycling protocol}. The absence of mirror-like symmetry of \textit{Segment 1} and \textit{Segment 3} together with the reduced polarization values of \textit{Segment 4} (compared with \textit{Segment 1}) explicitly relate to the slow decay and the irreversibility of $P_{induced}$. Our protocol, thus, rules out the contribution of any such irreversible $I_{P}$ contribution that may arise from release of $TSC$ or trapped charges at the vicinity of electrodes or due to thermally induced dipole relaxation mechanism. Other protocols of differentiating the $P_{S}$ from $P_{induced}$ pose their own experimental challenges. For instance, the method of sinusoidal temperature modulation requires creation of an undistorted pure sine wave, which is challenging. In contrast, \textit{4-segment thermal cycling protocol} is expedient due to the ease of setting up a constant temperature ramp rate. In particular, we employ the \textit{4-segment thermal cycling protocol} successfully to test the behavior of $P_{S}$ and $P_{induced}$ experimentally on standard prototype FE and multiferroic material along with other materials of interest, such as, HoCrO$_3$ and DyFe$_{0.5}$Mn$_{0.5}$O$_3$.

However, a cautious approach is advised while performing the \textit{4-segment thermal cycling protocol}: (1) Notably, the reversibility of the net (measured) $P_{S}$ requires a single-domain configuration. Hence, pre-poling of the sample is a prerequisite. (2) As the signal $I_{P_{S}} \propto \partial T / \partial t$, reasonable thermal ramp-rates can help to maximize the signal ($I_{P_{S}}$) compared to the background noise. Nevertheless, smaller rates of cooling/heating are favorable to ensure temperature-uniformity across the samples. (3) It is advisable to avoid crossing of any phase-transition other than the ferroelectric phase transition at $T_{C}$ while performing the \textit{4-segment thermal cycling protocol}. For instance, a first-order transition can pose difficulties in maintaining the constant ramp-rate due to the latent heat associated with the transition.

\section{Sample details and experimental set-up}

Glycine Phosphite (GPI) single crystal was grown using solution growth technique and a \textit{b}-plate was used for pyroelectric studies. Similarly, a single crystal of ME multiferroic GdMnO$_3$ was grown in-house with optical float-zone method \cite{Wagh2015}. Additionally, polycrystalline systems, HoCrO$_3$ and DyFe$_{0.5}$Mn$_{0.5}$O$_3$ were synthesized using spark plasma sintering technique (at 800$^{\circ}$C) \cite{Anusree2020} and conventional solid-state sintering (at 1150$^{\circ}$C), respectively. The $I_{P}$ measurements were performed on different pellets of these systems. Typically, silver paint electrodes were used on either side of the pellet to mimic the parallel plate capacitor geometry. In PE phase, a poling electric field was applied with the help of Keithley 2636 SMU (It is known that the poling procedure on polycrystalline materials can result in the net spontaneous polarization \cite{damjanovic1998ferroelectric}). The low temperature measurements were carried at certain ramp rates in Advanced Research Systems, Inc. Closed Cycle Cryostat. The $I_{P}$ current were recorded using Keithley Electrometer 6514A.

\section{Results and Discussion}

\subsection{Ferroelectric Glycine Phosphite and Magnetoelectric Multiferroic GdMnO$_3$}
\label{GPI and GMO}

\begin{figure*}
\centerline{\includegraphics[width=17cm,keepaspectratio]{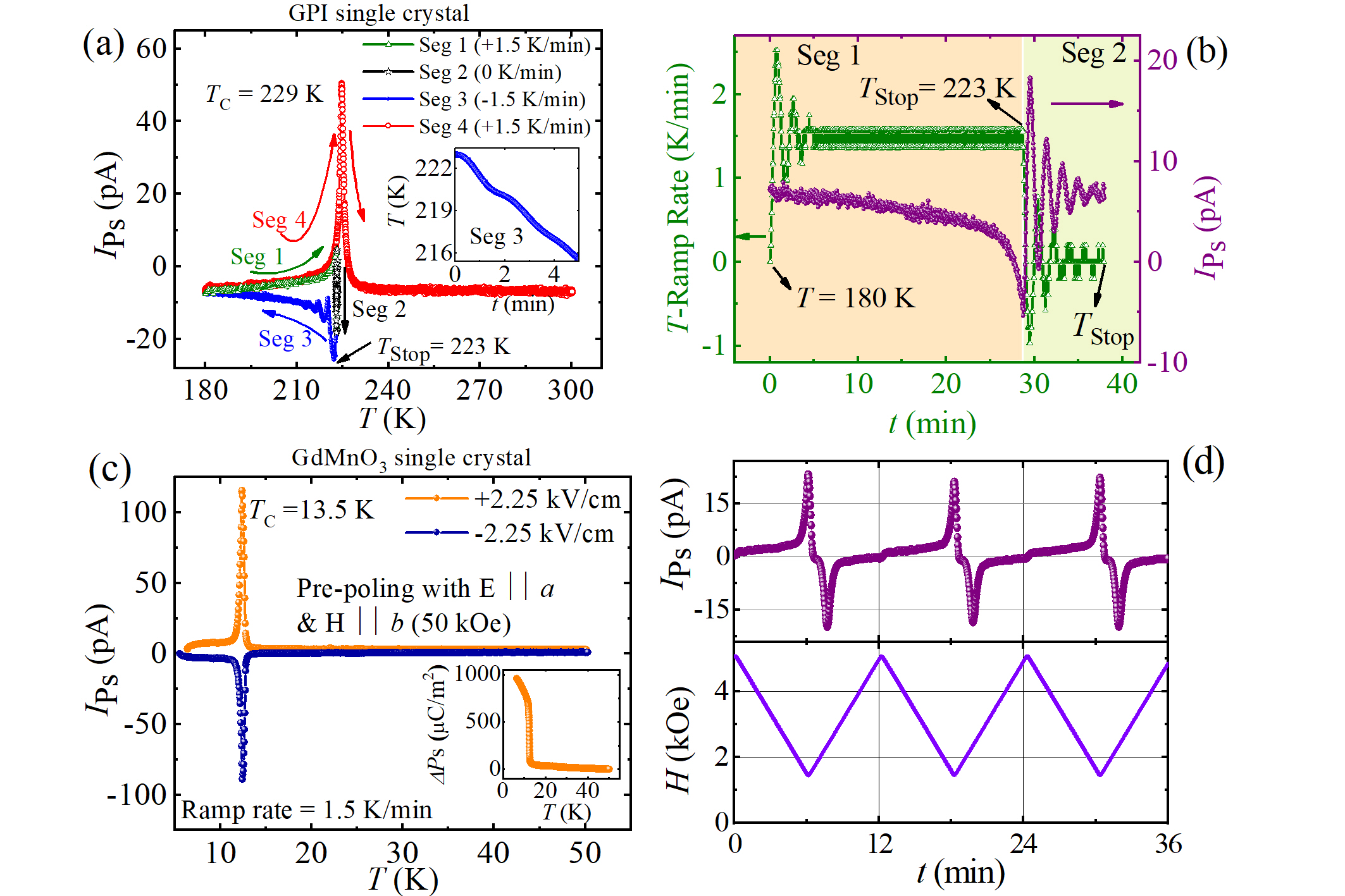}}
\caption{(a) \textit{4-segment thermal cycling protocol} of $I_{P_{S}}$ measurement in a single crystal GPI. Individual segments with specific ramp rates are marked accordingly. (b) The relation of $I_{P_{S}}$ and $T$ ramp rate as a function of $t$ in \textit{Segment 1} and \textit{Segment 2}. (c) Standard $I_{P_{S}}$ measurement protocol in GdMnO$_3$ single crystal with poling electric field $\pm $ 2.25 kV/cm applied along \textit{a}-axis, and $H$ = 50 kOe applied along \textit{b}-axis. Inset to figure is the estimated $P_{S}$ from the forward poling protocol of $I_{P_{S}}$. (d) The $H$-cycling as a function of $t$ along with the resultant change in $I_{P_{S}}$ measurement, recorded at ferroelectric state of GdMnO$_3$ at 8 K.}
\label{fig3}
\end{figure*}

We start discussion with polarization in single crystals of two prototype materials; a FE material, GPI, and a ME multiferroic material, GdMnO$_3$. GPI possesses monoclinic crystal structure. An anisotropy in the direction of \textit{b}-axis in GPI aligns $P_{S}$ along the same axis. FE ordering sets in at about $T_{C}$ = 233 K, and a sharp peak appears in the dielectric constant data in the \textit{b}-plate (not shown). The \textit{b}-plate was poled with $E_{P}$ = 3.2 kV/cm at $T_{P}$ = 300 K and cooled down to $T_{0}$ = 180 K. We employed the \textit{4-segment thermal cycling protocol}, as described in Section \ref{pyro}, with $b$ = 1.5 K, $T_{Stop}$ = 223 K, $Time_{Halt}$ = 30 min and $T_{End}$ = 300 K, and $I_{P_{S}}$ measured in all four segments as shown in Fig. \ref{fig3}(a).

In \textit{segment 1}, $I_{P_{S}}$ starts increasing rapidly as temperature approaches $T_{Stop}$. 
Fig. \ref{fig3}(b) illustrates the time variation of $b$ and $I_{P_{S}}$ in \textit{segment 1} and \textit{Segment 2}. At the beginning of \textit{Segment 1}, $b$ is 0 K/min. After oscillating for a while, the value of $b$ stabilizes at the set value of 1.5 K/min. The choice of temperature controlling PID values resulted in this decaying oscillations around the set value of $b$. Similar decaying oscillations are observed when the value of $b$ is set to zero at the beginning of the \textit{Segment 2}. As $I_{Ps} \propto \partial T / \partial t$ (see Eq. (\ref{eq:2})), the sign and the magnitude of $I_{P_{S}}$ depend on those of $b$. Hence, we expect $I_{P_{S}}$ must oscillate with $b$. At the start of the \textit{Segment 1}, $I_{P_{S}}$ oscillates, but the signal is feeble and almost concealed by the background noise in the measurement. This is mainly due to the small values of $\partial P_{S}/\partial T$ near $T_{0}$. On the other hand, we see the decaying oscillations clearly in $I_{P_{S}}$ at the start of \textit{Segment 2}, which are in phase with $b$ as expected. As soon as $b$ truly decays to zero in \textit{Segment 2}, $I_{P_{S}}$ falls to zero and, further, remains constant for the remaining duration of $Time_{Halt}$. A curve marked \textquoteleft Seg 2' represents the entire \textit{Segment 2} in Fig. \ref{fig3}(a).

In \textit{Segment 3}, only the sign of $I_{P_{S}}$ reversed while the magnitude remained unchanged during the cooling cycle with $-b$. As a consequence, a mirror-like symmetric curve is obtained as compared to \textit{Segment 1}. Here again, a few decaying oscillations of $I_{P_{S}}$ are observed at the start of the \textit{Segment 3}, before the magnitude of $-b$ stabilizes (see inset in Fig. \ref{fig3}(a)). In poled single-domain samples, the characteristic mirror-like symmetric relation between the \textit{Segment 1} and the \textit{Segment 3} results from the reversible nature of net $P_{S}$ with $T$. During the warming cycle, the $I_{P_{S}}$ values in \textit{Segment 4} coincide with those in \textit{Segment 1} before reaching $T_{Stop}$. Later, it increases and shows a sharp peak before dropping to zero at $T_{C}$. Reverse poling of the $b$-plate results in the $I_{P}$ sign reversal (not shown). The $I_{P}$ measurements performed in GPI single crystal by following the \textit{4-segment thermal cycling protocol} (see Fig. \ref{fig3}(a)) are consistent with simulated data of FE material (see Fig. \ref{fig1}(d)).

Thus, using the \textit{4-segment thermal cycling protocol}, the two important features of FE namely, highly responsive nature of $I_{P_{S}}$ to changes in the sign and the magnitude of $b$, and the reversible nature of the net $P_{S}$ with $T$ are confirmed.

In ME multifferroics, $P_{S}$ can be manipulated with $H$ and the corresponding ME currents can be distinguished from depolarizing currents originating from $P_{Induced}$ using magnetic-field-cycling similar to the \textit{4-segment thermal cycling protocol}. We discuss here a case of ME multiferroic GdMnO$_3$, possessing orthorhombic crystal structure. It exhibits paramagnetic to incommensurate collinear antiferromagnetic (AFM) phase transition at 42 K and subsequent canted AFM phase transition near 23 K. On the application of $H$ (above 10 kOe) along the \textit{b}-axis, the canted AFM phase transforms to cycloidal magnetic phase which is FE in nature. 

We poled the GdMnO$_3$ single crystal with $E_{P}$ = 2.25 kV/cm along the \textit{a}-axis at $T_{P}$ = 60 K in presence of $H$ of 50 kOe along the \textit{b}-axis. After cooling the sample to $T_{0}$ = 5 K, the electric field alone was removed. For both the cases of forward and reverse poling, the $I_{P_{S}}$ values were recorded under 50 kOe and $b$ = 1.5 K/min during the warming cycle as shown in the data plots in Fig. \ref{fig3}(c). A sharp peak of $I_{P_{S}}$ is visible at 13.5 K; its value falls to zero at about $T_{C}$ = 15 K at 50 kOe. $\Delta P_{S}$ for forward poling was estimated by integrating $I_{P_{S}}$ with time and shown in an inset of the Fig. \ref{fig3}(c). To analyze the response of $I_{P_{S}}$ and $P_{S}$ to changes in $H$ at 8 K, we cycled the field from 50 kOe to 1.5 kOe at a ramp rate of $\pm 100$ Oe/s (see Fig. \ref{fig3}(d)). It is evident that the $I_{P_{S}}$ curve inverts to a mirror-like symmetric curve with the reversal of the ramp rate. This indicates reversibility of the $P_{S}$ with $H$. As the ramp rate reverses through zero, it dwells there for finite time during which the $I_{P_{S}}$ value becomes zero momentarily. In the ME multiferroic under investigation, the time scales involved in the change in $P_{S}$ with $H$ are small when compared with experimental time scale, under which practical assumption of $I_{P_{S}} \propto \partial P_{S} / \partial H$ holds good.

\subsection{Simple Perovskite HoCrO$_3$ and B-site Disordered Perovskite DyFe$_{0.5}$Mn$_{0.5}$O$_3$}
\label{HCO and DFMO}

We shall now focus to systems such as HoCrO$_3$ (HCO), a material belonging to simple perovskite family of rare-earth chromates and DyFe$_{0.5}$Mn$_{0.5}$O$_3$ (DFMO), member of B-site disordered perovskite family. Both the family of materials are constantly under scrutiny for FE and multiferroicity. In this regard, our study represents the charaterization of $I_{P}$ measurements on members of such families, HCO and DFMO using our \textit{4-segment thermal cycling protocol} as follows. 

At first, we investigate the behavior of HCO wherein, the FE order was reported earlier \cite{Ghosh2015}, based on the standard $I_{P}$ measurement protocols using forward and reverse poling. 
FE order is perceived to set in at a higher temperature of 240 K which is different than the magnetic ordering temperature. We investigate the nature of the FE order (if present) in HCO using using our thermal cycling protocol.

\begin{figure*}
\centerline{\includegraphics[width=17cm,keepaspectratio]{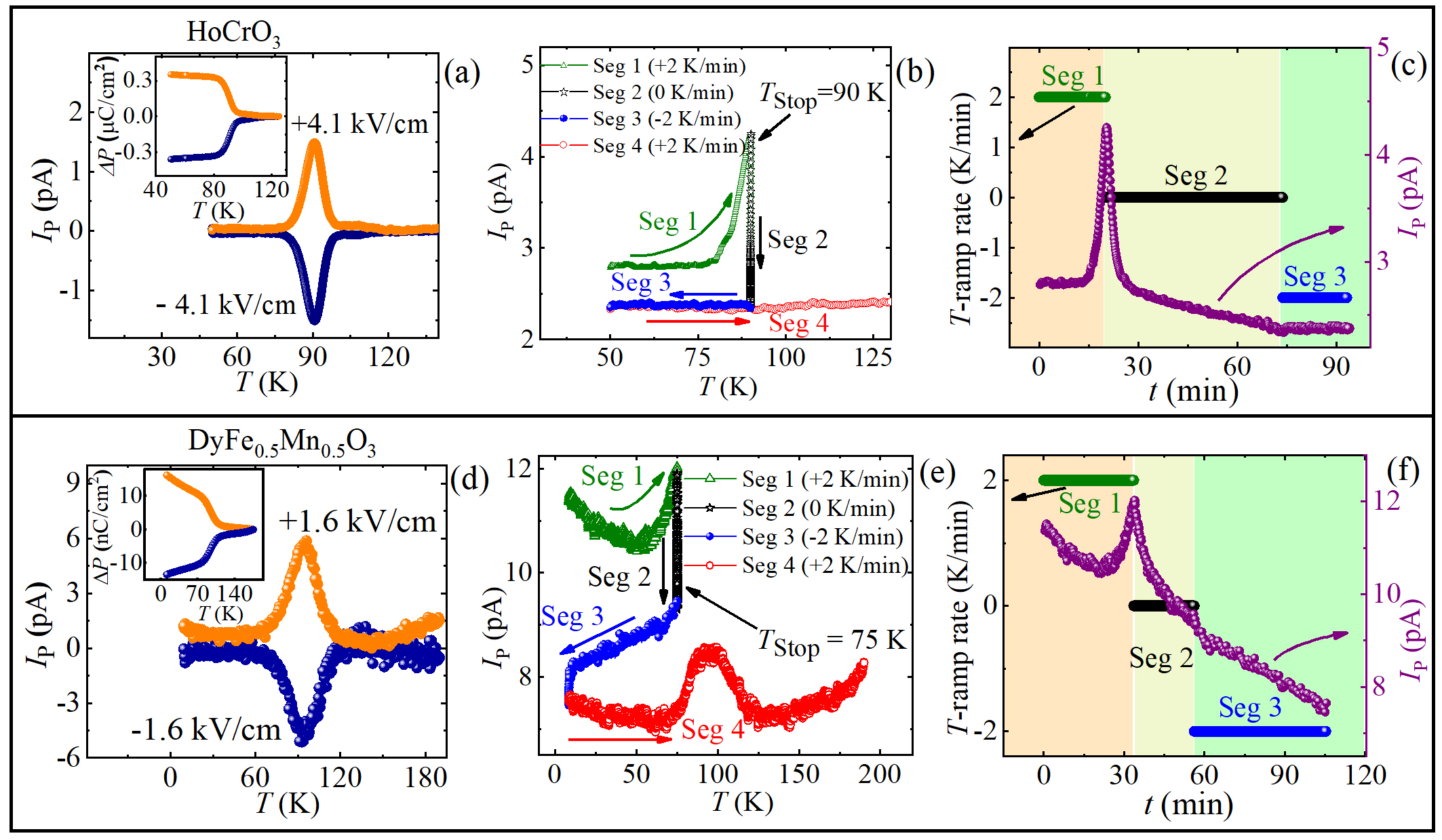}}
\caption{$I_{P}$ measurements on HoCrO$_3$ (HCO) ((a) to (c)) and DyFe$_{0.5}$Mn$_{0.5}$O$_3$ (DFMO) ((d) to (f)) pellets. (a) and (d) are $I_{P}$ vs. $T$ curves (forward and reverse poling protocols) for HCO and DFMO, respectively. Insets of these plots show temperature variation of estimated $\Delta P$. (b) and (e) are $I_{P}$ measurements in response to the \textit{4-segment thermal cycling protocol} in HCO and DFMO, respectively. (c) and (f) are the plots of $I_{P}$ and $T$ ramp rates as a function of $t$ for HCO and DFMO, respectively. (Figure (a) and (b) utilizes the data from our earlier work \cite{Anusree2020}).}
\label{fig4}
\end{figure*}

HCO is orthorhombic at room temperature and exhibits antiferromagnetic insulating behavior with N\'eel temperature, $T_{N}$ = 140 K \cite{Anusree2020}. We performed $I_{P}$ measurement on an HCO polycrystalline pellet of 8 mm diameter and 0.35 mm thickness. Poling was performed with $E_{P}$ of $\pm 4.1$ kV/cm at $T_{P}$ of 300 K and the sample was later cooled down to $T_{0}$ = 50 K. Standard $I_{P}$ measurements were initially carried out while warming the HCO pellet with forward and reverse poling protocols from $T_{0}$ to a $T_{End}$ of 300 K. Figure \ref{fig4}(a) shows the $I_{P}$ in the range of 50 K to 125 K (data for $T >$ 125 K will be discussed later in this Section). The inset of Fig. \ref{fig4}(a) exhibits the plot of $\Delta P$ as a function of $T$ estimated by integrating the $I_{P}$ values in the above range of 50 K to 125 K over the time scale of the measurement. A relatively sharp peak in $I_{P}$ is observed at 91 K, and as expected, the $I_{P}$ switches its sign for the reverse poling scenario. This peak is highly reproducible and tested for different $E_{P}$ (not shown). Estimate of $\Delta P$ from $I_{P}$ in the range of 50 K to 125 K gives non-zero values of $\Delta P$ for $T <$ 125 K. It may be noted that this temperature is different from the $T_{N}$ of HCO.

Further, to validate the estimated $\Delta P$ corresponding to the peak at 91 K, and to rule out extraneous effects related to non-pyroelectric current, we used the \textit{4-segment thermal cycling protocol} with $b$ = 2 K/min, $T_{Stop}$ = 90 K, $Time_{Halt}$ = 60 min and $T_{End}$ = 300 K. The data are plotted in Fig. \ref{fig4}(b). The corresponding plots of $b$ and $I_{P}$ as a function of time elapsed for different segments of the protocol are shown in Fig. \ref{fig4}(c). At \textit{Segment 1}, the $I_{P}$ curve starts rising for a chosen value of $b$, in the vicinity of 91 K peak. However, the system is halted at $T_{Stop} <$ 91 K in \textit{Segment 2} with $b$ set to 0 K/min. At this point, $I_{P}$ falls exponentially in \textit{Segment 2} as in Fig. \ref{fig4}(c)) for elapsed $Time_{Halt}$. It is interesting to note that $I_{P}$ value does not reach zero instantly in \textit{Segment 2} when $b$ is set to zero, in contrast to the highly responsive behavior of $I_{P_{S}}$ in the case of GPI and GdMnO$_3$. In the case of HCO, $I_{P}$ decreases to the background limit only at the end of \textit{Segment 2} (shown in Fig. \ref{fig4}(c)) exhibiting much slower response to the change in $b$ when compared to GPI and GdMnO$_3$.

In \textit{Segment 3} with $-b$, $I_{P}$ continues to remain in the background as seen from Fig. \ref{fig4}(b) and (c). The absence of mirror-symmetry between \textit{Segment 1} and \textit{Segment 3} demonstrates the irreversible nature of $\Delta P$ in HCO (unlike that of $P_{S}$ in GPI). In continuing with thermal cycling at the beginning of \textit{Segment 4}, the value of $I_{P}$ is found to be different from that of \textit{Segment 1}. Hence, the estimated differed values of $\Delta P$ from these $I_{P}$ values provide evidence for excluding $P_{S}$ in HCO. Even, the $I_{P}$ peak at 91 K has disappeared due to the fully depolarized $\Delta P$ in \textit{Segment 4}. Hence, our experiments indicate the absence of $P_{S}$ in HCO. 

It is worthwhile to mention the resemblance of the \textit{4-segment thermal cycling protocol} in HCO to the faster mechanism of dipolar relaxation phenomenon (Section \ref{TSC} (Fig. \ref{fig2}(b))) in reference to $TSC$ and $P_{Induced}$. The nature of depolarization of $\Delta P$ in HCO around the 91 K peak in $I_{P}$ can be correlated very well to the $TSC$ mechanism. This indicates the presence of $P_{Induced}$ alone and thus, the absence of $P_{S}$ in the range of 50 K to 125 K in HCO. 

\begin{figure}
\centerline{\includegraphics[width=7cm,keepaspectratio]{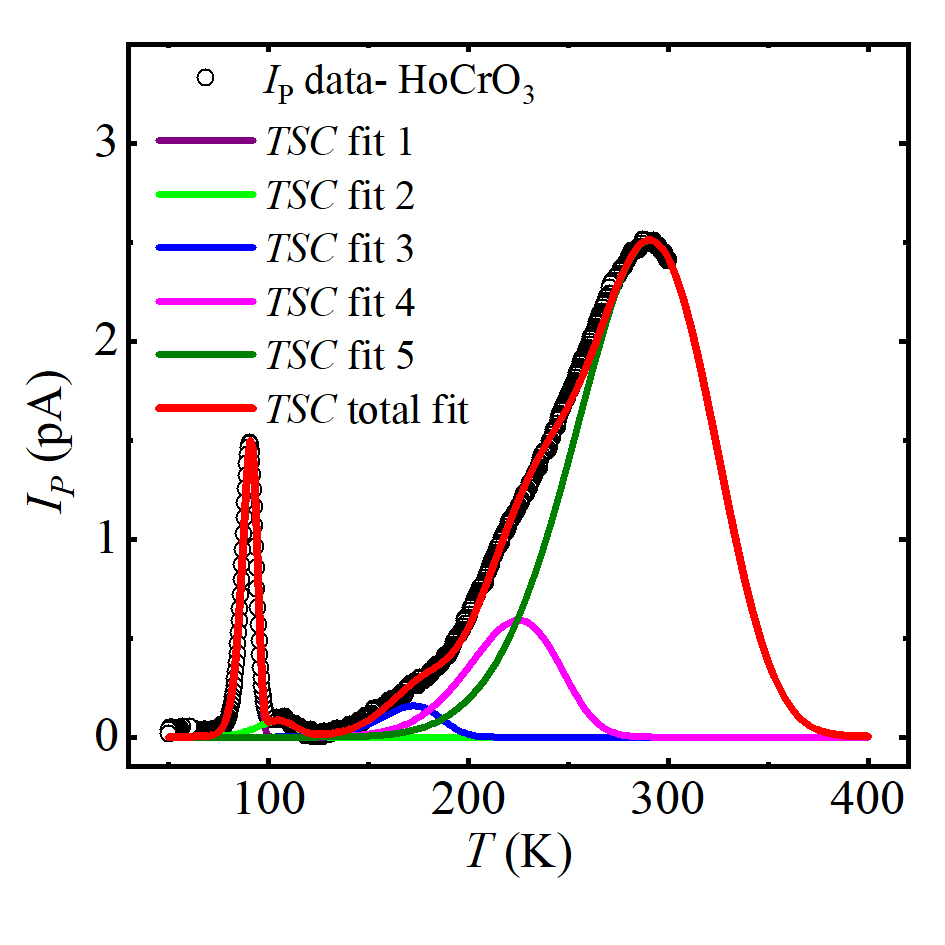}}
\caption{Standard $I_{P}$ measurements on HoCrO$_3$ for the forward poling scenario from 50 K to 300 K. The de-convolution of the experimental data of HoCrO$_3$ into a set of thermally stimulated current peaks comprising various activation energies and time constants (see Table \ref{tab:table1}).}
\label{fig5}
\end{figure}

A comment can be made here on the observed sharp $TSC$ peak at 91 K in HCO material by drawing the results of our earlier Synchrotron X-ray diffraction study on HCO \cite{Anusree2020}, which revealed a structural distortion at 100 K. In our earlier study \cite{Anusree2020}, we hypothesized that faster depolarization of $P_{Induced}$ in the vicinity of a structural distortion suddenly releases space charges which give rise to the $TSC$ peak in $I_{P}$ data. However, elaborate experimental evidence is needed to clarify the role of structural distortion in $TSC$. Further, to understand the origin of this peak, we have quantitatively analysed the $I_{P}$ measured over a wide range of temperatures up to 300 K (see Fig. \ref{fig5}). This is explained in detail in the latter part of this section. 

The B-site disordered orthochomites have been previously the subject of extensive studies for its FE ordering \citep{yuan2016observation,rajeswaran2012ferroelectricity}. As a corollary, the less explored DFMO is a good prospect due to its antiferromagnetic insulating characteristics and $T_{N} \approx$ 320 K \cite{chiang2011effect}. In this context, a detailed study of $I_{P}$ measurements using the \textit{4-segment thermal cycling protocol} on DFMO is in order for which an $I_{P}$ measurement protocol used for HCO was adapted. Polycrystalline pellet of 8 mm diameter and 1 mm thickness was used for the purpose. The results obtained thereof are given in Fig. \ref{fig4}(d) to (f). DFMO exhibits increase in resistivity as temperature is lowered. Hence, we chose $T_{P}$ = 190 K where, the resistivity of the sample is reasonably large for applying $E_{P}$ = $\pm 1.6$ kV/cm. Standard $I_{P}$ measurements were carried out while warming the sample from $T_{0}$ (= 8 K) to $T_{End}$ (= 200 K) with $b$ = 2 K/min (See Fig. \ref{fig4}(d)). We observed a broad peak near 95 K in both the forward and reverse poling scenario. In order to investigate the nature of the broad peak at 95 K, we performed the \textit{4-segment thermal cycling protocol} on DFMO by choosing the $T_{Stop}$ to be at 88 K with a $Time_{Halt}$ of 30 min (Refer Fig. \ref{fig4}(e) and (f)). Notably, during $Time_{Halt}$ in \textit{Segment 2} the recorded current did not instantly drop to zero but remained finite and exhibited a gradual decay. This depolarization process is seen to be incomplete at the end of \textit{Segment 2} and the current continues to fall further to lower values in \textit{Segment 3} as well (Refer Fig. \ref{fig4}(e) and (f)). This fall of $I_{P}$ continues further at the beginning of \textit{Segment 4}. Here, a peak with smaller intensity reappeared in a new position at 88 K. Notable observations in DFMO experiments (including a few similarities with that of HCO) include; (1) the absence of mirror symmetry between \textit{Segment 1} and \textit{Segment 3} (2) slower response of $I_{P}$ to the changes in $b$ (3) dissimilar $I_{P}$ values  at the beginning of \textit{Segment 1} and \textit{Segment 4} (4) shifted peak position of $I_{P}$ in \textit{Segment 4}. From all the above points, we can conclude that the 95 K peak in DFMO is not synonymous with FE nature since those highly responsive and reversible attributes of $P_{S}$ are lacking.

Citing the similarities in the $I_{P}$ data of DFMO and the $TSC$ mechanisms discussed in Section \ref{TSC}, the resemblance with a slow depolarization mechanism of $P_{Induced}$ (Fig. \ref{fig2}(a)) yields more credence to the $TSC$ mechanisms in place for DFMO. In addition, the dependency of $b$ on $TSC$ peak position was also tested in case of DFMO with $b$ = 3 K/min (not shown). The $I_{P}$ peak shifts to 105 K for $b$ = 3 K/min, a result that validates the simulation data presented in Fig. \ref{appendixfig1}(b) of Appendix. It provides conclusive evidence that the origin of the peak in DFMO is traceable to $TSC$ and not to FE ordering.

As mentioned earlier, Fig. \ref{fig5} shows the $I_{P}$ curve measured for HCO up to $T_{P}$ = 300 K for the forward poling scenario. It is discernible that above 125 K, $I_{P}$ again increases and shows a very broad feature peaking just below 300 K. We tried fitting sharp as well as broad features along with a few observed shoulder-features with five independent $TSC$ peaks. The fitting parameters such as the peak temperature in $TSC$ ($T_{M}$), the initial value of $P_{Induced}$ at the beginning of the warming ($P_{0}$), the activation energy ($E_{a}$) and the time constant ($\tau_{0}$) are listed in the Table \ref{tab:table1}. The total fit, a combined effect of all the five peaks, is denoted by a red curve and matches well with the experimental data. The sharp peak near 91 K is accounted by the $TSC$ Fit 1 (see Fig. \ref{fig5}) having a relatively high $E_{a} = 189.0(8)$ meV and a low $\tau_{0} = 3.7(4)$ ns and in turn, it describes the faster depolarization observed in the narrow temperature window. Whereas, the $\tau_{0}$ values, corresponding to other $TSC$ fits ($TSC$ Fit 2 to 5) with higher $T_{M}$, are relatively high and falls in the range, 10 to 700 ms. Also, magnitudes of $E_{a}$ are found to increase systematically with $T_{M}$ in $TSC$ Fits, 2 to 5. Our analysis confirms that the observed peaks in $I_{P}$ in HCO are a manifestation of $TSC$ due to depolarization of $P_{Induced}$ during warming. However, the physical reasons for the observation of different $TSC$ peaks of various  $E_{a}$ is not covered in this study. 

In essence, it is readily perceptible that the standard $I_{P}$ measurement protocols using forward and reverse poling alone are not sufficient to comment on the presence of $P_{S}$ in a material under scrutiny (HCO and DFMO, in our case). The switching of the sign of $\Delta P$ with the reverse poling scenario can hold good for both $P_{S}$ and $P_{Induced}$. Hence, a careful approach is recommended while performing the $I_{P}$ measurements in samples comprising either $P_{Induced}$ alone or, those with both $P_{S}$ and $P_{Induced}$ coexisting. In this regard, adapting the \textit{4-segment thermal cycling protocol} in $I_{P}$ measurement will readily assist in distinguishing the $P_{S}$ from $P_{Induced}$, and thus avoid inconspicuous and erroneous conclusions of FE ordering and multiferroicity.

\begin{table}
\caption{\label{tab:table1}%
De-convolution of measured broad current peak in \ch{HoCrO_{3}} in to various contributing individual thermally stimulated current ($TSC$) peaks and associated parameters.
}
\begin{ruledtabular}
\begin{tabular}{lcccc}
\textrm{$TSC$ Peak\footnote{De-convoluted peaks.}}&
\textrm{$T_{M}$(K)\footnote{Temperature at which $TSC$ exhibits a peak value.}}&
\multicolumn{1}{c}{\textrm{$P_{0}$(C)\footnote{$P_{Induced}$ in the sample at temperature $T_{0}$ before starting the thermal ramp run.}}}&
\multicolumn{1}{c}{\textrm{$E_{a}$(meV)\footnote{Activation energy associated with the mechanism of $P_{Induced}$.}}}&
\multicolumn{1}{c}{\textrm{$\tau_{0}$(s)\footnote{Time constant associated with the mechanism of $P_{Induced}$ as $T\to\infty$.}}}\cr
\colrule
$TSC$ Fit 1 & 91  & $1.0(1) \times 10^{-1}$ & 189.0(8) & $3.7(4) \times 10^{-9}$\\
$TSC$ Fit 2 & 105 & $5(3) \times 10^{-9}$ & 90(5) & $1.5(3) \times 10^{-2}$\\
$TSC$ Fit 3 & 173 & $11(2) \times 10^{-9}$ & 154(3) & $1.3(8) \times 10^{-2}$\\
$TSC$ Fit 4 & 225 & $10.1(9) \times 10^{-9}$ & 173(2) & $0.9(1) \times 10^{-1}$\\
$TSC$ Fit 5 & 291 & $9.5(7) \times 10^{-9}$ & 184(2) & $6.9(6) \times 10^{-1}$\\
\end{tabular}
\end{ruledtabular}
\end{table}

\section{Conclusion}
In summary, we have proposed a simple modified protocol for pyroelectric current measurements termed as, \textit{4-segment thermal cycling protocol}. A clear and distinct behavior of irreversible induced polarization ($P_{Induced}$) can accurately be discerned from that of a reversible and spontaneous electric polarization ($P_{S}$), with the help of above mentioned protocol. Thus, our systematic approach primarily helps to avoid the possibilities in misinterpretation of the ferroelectric (FE) phase in a given material. At first, we simulated current response of an FE material consisting of pure $P_{S}$, and a non-FE material comprising only $P_{Induced}$ to the above protocol. Further, these thermal cycling protocols are experimentally demonstrated on a single crystal of a prototype FE system, Glycine Phosphite. Besides, some equivalence between our thermal cycling protocol and the typical magnetic field cycling method used for characterization of magnetoelectric (ME) multiferroics  was discussed with measurements on ME GdMnO$_3$ single crystal. Later, we have employed these protocols to investigate two polycrystalline materials, HoCrO$_3$ and DyFe$_{0.5}$Mn$_{0.5}$O$_3$ where HoCrO$_3$ has been reported to be multiferroic elsewhere. Our studies highlight dominant contribution of thermally stimulated currents ($TSC$) in pyroelectric measurements in both the materials. Quantitative analysis in HoCrO$_3$ system based on an existing theory of $TSC$, allowed us to estimate various activation energies and time constants associated with different depolarization mechanisms. In conclusion, our \textit{4-segment-thermal cycling protocol} establishes that HoCrO$_3$ and DyFe$_{0.5}$Mn$_{0.5}$O$_3$ do not possess any reversible and spontaneous ferroelectric polarization and thus are not multiferroic in nature within the temperature range tested.

\section*{ACKNOWLEDGEMENTS}
S.G.B. acknowledges DST INSPIRE Faculty Fellowship for the financial assistance. P.S.A.K. thanks Nano Mission, Department of Science and Technology, India.

A.A.W. and S.G.B. contributed equally to this work.


\appendix*

\section{Theory of Thermally Stimulated Currents}
\label{appendix}
This section describes depolarization mechanism of induced polarization responsible for thermally stimulated currents ($TSC$) when materials are warmed (or cooled) at constant ramp rates. In this section, we consider dipolar relaxation phenomena \cite{Garn1982,Perlman1971} and assume that the polarization ($P$) can be induced in the material at constant temperature by applying an electric field ($E$). Subsequent removal of $E$ results in depolarization and the time rate of depolarization can be described by the following differential equation wherein $\tau$ is an associated time constant.
\begin{equation}
\frac{dP(t)}{dt} = \frac{-P(t)}{\tau}
\label{eq:diffeqn}
\end{equation}

Solving Eq. (\ref{eq:diffeqn}) for $P(t)$ gives,

\begin{subequations}
\label{eq:whole}
\begin{equation}
P(t) = P_{0} e^{-\frac{t}{\tau}}
\label{eq:2a}
\end{equation}
where, $P_{0}$ is an initial value of polarization.
\begin{equation}
J_{TS}(t) = -\frac{P_{0}}{\tau }  e^{-\frac{t}{\tau }}
\label{eq:2b}
\end{equation}
\end{subequations}

\begin{figure}
\centerline{\includegraphics[width=7cm,keepaspectratio]{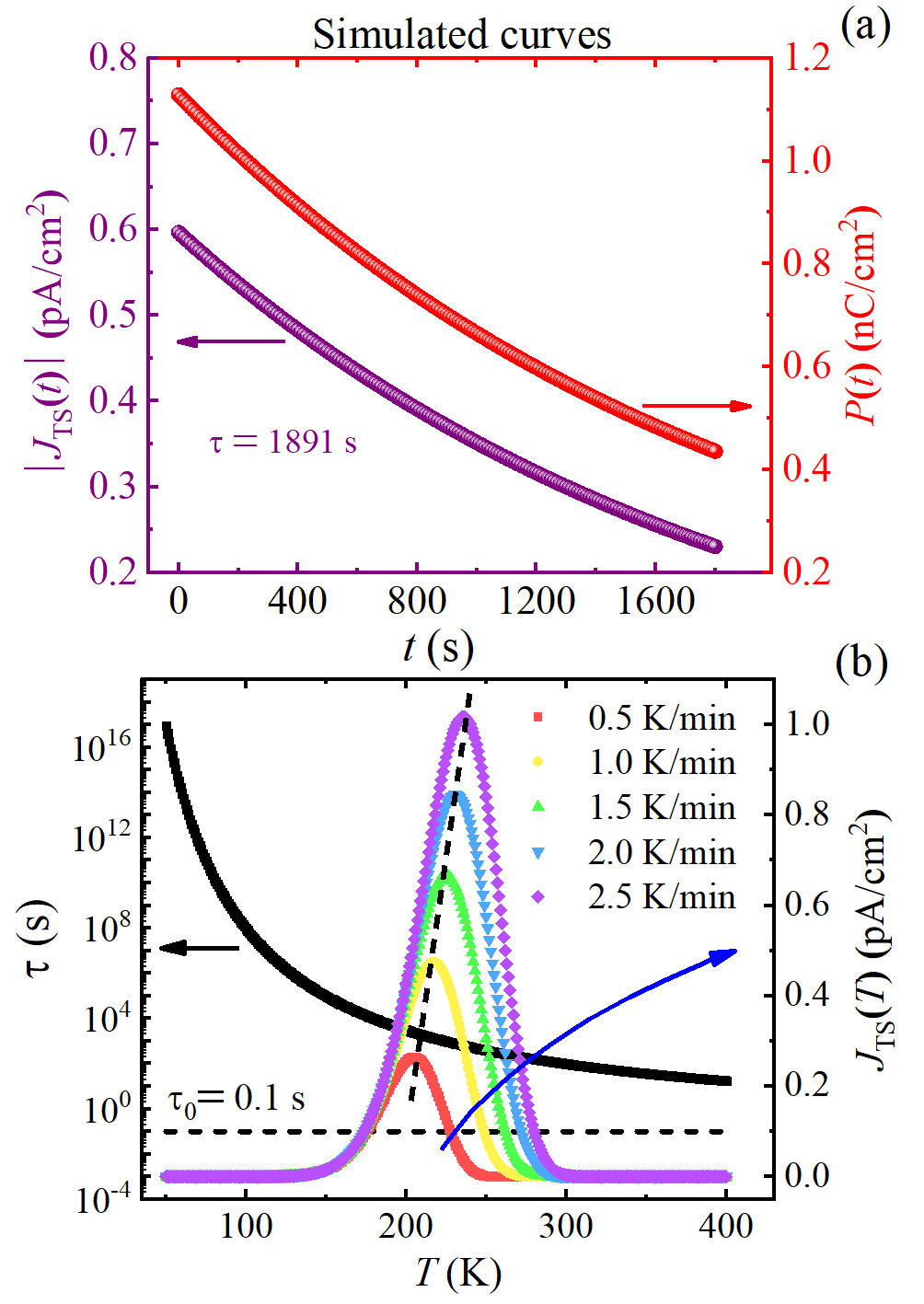}}
\caption{(a) Time evolution of depolarization ($P_{t}$) and depolarization current density ($J_{TS}(t)$), and (b) temperature variation of time constant and effect of warming ramp-rates on the peak-position in $J_{TS}(T)$.}
\label{appendixfig1}
\end{figure}

Depolarization current is evaluated in Eq. (\ref{eq:2b}) by differentiating Eq. (\ref{eq:2a}) with respect to time. An exponential decrease in both, $P(t)$ and the depolarization current density ($J_{TS}(t)$), with time are plotted in Fig. \ref{appendixfig1}(a) for $\tau$ = 1891 s. The $\tau$
 depends on temperature and increases at lower temperature as,
\begin{equation}
\tau = \tau_{0} e^{\frac{E_{a}}{k T}}
\label{eq:A3}
\end{equation}
where, $E_{a}$ is an activation energy related to depolarization mechanism, $k$ is the Boltzmann constant and $\tau_{0}$ is a magnitude of the $\tau$ as $T\to\infty$.  A generic behaviour of $\tau$ with temperature, as governed by Eq. (\ref{eq:A3}), is shown in Fig. \ref{appendixfig1}(b) for values of $\tau_{0}$ and $E_{a}$ to be 0.1 s and 0.1782 eV, respectively. Notably, $\tau$ attains very high values ($\sim 10^{16}$ s) at lower temperature. Substituting Eq. (\ref{eq:A3}) in Eq. (\ref{eq:diffeqn}) and solving it for $P(t)$ gives the time evolution of depolarization at any constant temperature $T$,
\begin{equation}
P(t) = P_{0} e^{-\frac{t e^{-\frac{E_{a}}{k T}}}{\tau_{0}}}.
\label{eq:A4}
\end{equation}
In particular, by employing a warming (or cooling) protocol with constant ramp rate ($b$) the temperature itself becomes a function of time i.e. $T(t)$ with
\begin{equation}
T(t) = T_{0} + b t
\label{eq:A5}
\end{equation}
where, $T_{0}$ is the initial temperature at $t = 0$. Substitution of Eq. (\ref{eq:A3}) and Eq. (\ref{eq:A5}) in Eq. (\ref{eq:diffeqn}) and further solving for $P(t)$ yields Eq. (\ref{eq:PdwithTime}).

\begin{widetext}

\begin{equation}
P(t) = P_{0} \exp \left(-\frac{E_{a} \text{Ei}\left(-\frac{E_{a}}{b k t+k T_{0}}\right)+k (b t+T_{0}) e^{-\frac{E_{a}}{b k t+k T_{0}}}-E_{a} \text{Ei}\left(-\frac{E_{a}}{k T_{0}}\right)+k T_{0} \left(-e^{-\frac{E_{a}}{k T_{0}}}\right)}{b k \tau_{0}}\right)
 \label{eq:PdwithTime}
\end{equation}

\begin{equation}
P(T) = P_{0} \exp \left(\frac{-E_{a} \text{Ei}\left(-\frac{E_{a}}{k T}\right)+E_{a} \text{Ei}\left(-\frac{E_{a}}{k T_{0}}\right)+k T \left(-e^{-\frac{E_{a}}{k T}}\right)+k T_{0} e^{-\frac{E_{a}}{k T_{0}}}}{b k \tau_{0}}\right)
 \label{eq:PdwithT}
\end{equation}

\begin{equation}
J_{TS}(t) = -\frac{P_{0}}{\tau_{0}} \exp \left(-\frac{\frac{E_{a} \left(\text{Ei}\left(-\frac{E_{a}}{b k t+k T_{0}}\right)-\text{Ei}\left(-\frac{E_{a}}{k T_{0}}\right)\right)}{k}+\frac{b E_{a} \tau_{0}}{b k t+k T_{0}}+b t e^{-\frac{E_{a}}{b k t+k T_{0}}}+T_{0} e^{-\frac{E_{a}}{b k t+k T_{0}}}-T_{0} e^{-\frac{E_{a}}{k T_{0}}}}{b \tau_{0}}\right)
 \label{eq:JdwithTime}
\end{equation}

\begin{equation}
J_{TS}(T) = -\frac{P_{0}}{\tau_{0}} \exp \left(-\frac{\frac{E_{a} \left(\text{Ei}\left(-\frac{E_{a}}{b k t+k T_{0}}\right)-\text{Ei}\left(-\frac{E_{a}}{k T_{0}}\right)\right)}{k}+\frac{b E_{a} \tau_{0}}{b k t+k T_{0}}+b t e^{-\frac{E_{a}}{b k t+k T_{0}}}+T_{0} e^{-\frac{E_{a}}{b k t+k T_{0}}}-T_{0} e^{-\frac{E_{a}}{k T_{0}}}}{b \tau_{0}}\right)
 \label{eq:JdwithT}
\end{equation}

\end{widetext}

where, \text{Ei} is an exponential Integral function. In order to get Eq. (\ref{eq:PdwithT}) for $P(T)$, the time ($t$) in Eq. (\ref{eq:PdwithTime}) is replaced with $(T-T_{0})/b$. $P(T)$ is the temperature evolution of depolarization for the thermal ramp starting from a temperature ($T_{0}$) at constant $b$. $J_{TS}(t)$ is computed by differentiating Eq. (\ref{eq:PdwithTime}) with respect to time, $t$ (see Eq. (\ref{eq:JdwithTime})). Further, the temperature evolution of  depolarization current density ($J_{TS}(T)$) for the thermal ramp starting from $T_{0}$ at constant $b$, is expressed in Eq. (\ref{eq:JdwithT}) after the replacement of $t$ with $(T-T_{0})/b$ in Eq. (\ref{eq:JdwithTime}).

Figure \ref{appendixfig1} conveys a behaviour of thermally stimulated currents in a nutshell. Particularly, $|TSC(T)|$ (i.e. $|J_{TS}(T)|$) curves are simulated using Eq. (\ref{eq:JdwithT}) and plotted in Fig. \ref{appendixfig1}(b). Here, we consider a typical electric poling process on a sample having no ferroelectricity. Hence, the poling results only in an induced polarization.  $P_{0}$ at $T_{0}$ is estimated as, $P_{0} = \frac{N \mu^{2} E_{P}}{3 k T_{P}}$ where, $E_{P}$ is an electric poling field, $T_{P}$ is the poling temperature, $\mu$ is a dipole moment and $N$ is a dipole concentration \cite{Perlman1971}.  With $P_{0} = 15 nC/cm^2$ at $T_{0}$ = 50 K and $\tau$ = 0.1 s, various $J_{TS}(T)$ curves corresponding to different ramp rates ($b$ = 0.5, 1.0, 1.5, 2.0, 2.5 K/min) are plotted for temperature range, 50 to 400 K. Notably, the peak position of $TSC$ shifts to higher temperature, with increase in the value of $b$.

In summary, warming (or cooling) of a non-ferroelectric sample with constant ramp rate, succeeding a typical electric poling process, can give a broad peak in the temperature evolution of thermally stimulated current. This peak needs to be distinguished from the peak that is typically observed in pyroelectric current measurement due to depolarization of spontaneous polarization in ferroelectric material near the transition temperature. Our protocol to distinguish between the two is discussed in the Section \ref{pyrosimulation} on page \pageref{pyrosimulation}.

\bibliography{PyroelectricStudies}

\end{document}